# Origin of dips in tunneling d$I$/d$V$ characteristics of cuprates

A. Mourachkine[*]

*Ecole Polytechnique Federal de Lausanne, Institut de Physique des Nanostructures, Batiment PH-B, Station 3, Lausanne CH-1015, Switzerland*

**Abstract**

Extensive efforts have been made to understand the electronic properties of high-$T_c$ superconductors. One feature which has been discussed in the literature during the past few years is the dips in tunneling conductances obtained in cuprates. In this contribution, we focus our attention on the origin of these dips. On the basis of experimental data obtained in $Bi_2Sr_2CaCu_2O_8$ and $YBa_2Cu_3O_7$, we show that the dips appear naturally in tunneling spectra due to a superposition of the peaks and humps and, therefore, have no physical meaning.

At present, 20 years after the discovery of high-$T_c$ superconductivity [1], there is no generally accepted theoretical model describing this remarkable phenomenon. So, the search for a clue continues: different experimental data obtained in cuprates are still widely discussed in the literature, including leading journals. Every feature in spectra of copper oxides is under the scrutiny. However sometimes, some features in spectra have nothing to do with superconductivity (SC), they reflect properties of a certain class of materials. As an example, a kink in energy dispersion obtained in angle-resolved photoemission (ARPES) measurements along the ($\pi$/2, $\pi$/2) direction in $n$-layered Bi-compounds ($n$ = 1-3) apparently has nothing to do with SC, contrary to the kink in dispersion in the ($\pi$, 0) direction [2]. Here we discuss the dips in tunneling conductances of cuprates.

Tunneling conductances obtained in cuprates at $T \ll T_c$ have the so-called peak-dip-hump structure. Quasiparticle (QP) peaks and humps at higher energies are separated by dips. This structure is more pronounced in spectra obtained in SC-insulator-SC (SIS) junctions; in SC-insulator-normal metal (SIN) junctions, the conductances are in general asymmetrical relatively zero bias. By analogy with phonon structures in conductances of conventional SCs, one group suggested in 1998 that the dips are caused by a bosonic excitation responsible for the pairing in cuprates [3]. This interpretation, still present in publications of the group, is erroneous because it contradicts a number of experimental facts. Due to limited space, we consider only a few of them.

In tunneling measurements performed by break junctions in heavily underdoped $Bi_2Sr_2CaCu_2O_{8+x}$ (Bi2212), the humps were observed below $T_c$ with *and* without QP peaks [4-6], as shown in Fig. 1. In Fig. 1(b), one can see at low bias the traces of undeveloped QP peaks[1]. In this case, it is obvious that the dips in conductances appear naturally due to a superposition of the peaks and the humps. As a consequence, the dips are meaningless.

ARPES measurements performed in Bi2212 fully support the latter scenario. Upon cooling, the QP peak appears in ARPES spectra above $T_c$ on a side of the hump [7]. So, the hump in ARPES spectra is a "background" for the peak [7]. Then, the dip between the peak and the hump has no physical meaning.

---

[*] *E-mail address:* andrei_mourachkine@yahoo.co.uk
[1] It is well known that SC in heavily underdoped region is "diluted"; this can be the reason for the absence of QP peaks in the conduactance in Fig. 1(b).

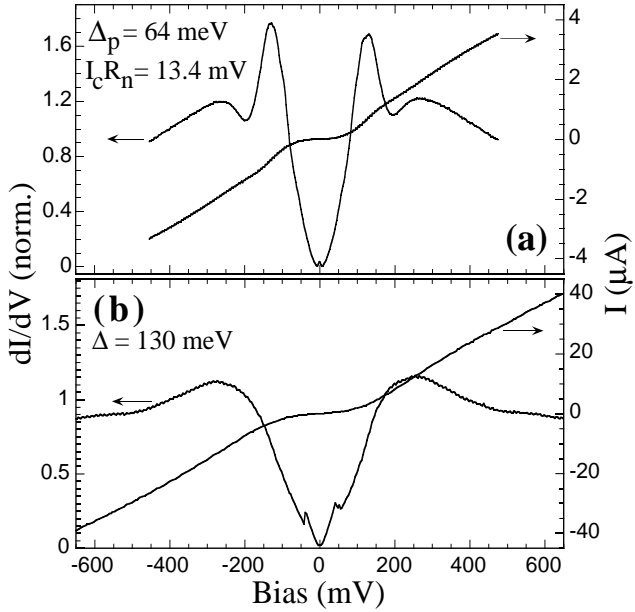

Fig. 1. (a) and (b) SIS tunneling $I(V)$ and $dI(V)/dV$ characteristics measured at 14 K within the same underdoped Bi2212 single crystal having $T_c = 51$ K. In both plots, the $dI(V)/dV$ are normalized at -400 mV. In plot (a), $I_cR_n$ denotes the Josephson product [4-6].

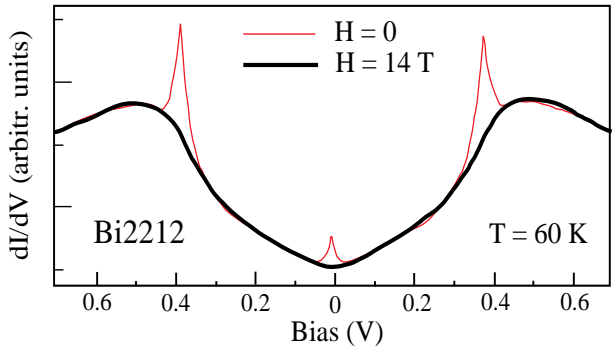

Fig. 2. Conductances for a Bi2212 mesa ($T_c = 92$ K) at $T = 60$ K for 0 T (thin line) and 14 T (thick line). The existance of normal-state "background" is clearly seen [8]. The field is applied along the $c$-axis.

Another way to demonstrate that the peaks appear on the inner sides of the humps is to suppress the SC state by magnetic field. The problem, however, is that the values of $B_{c2}(0)$ in hole-doped cuprates are incredibly high [6]. At the same time, at $T \sim T_c$ the $B_{c2}$ values necessary to suppress SC in cuprates are accessible in laboratory conditions. Figure 2 shows $c$-axis interlayer tunneling spectra obtained in Bi2212 at $T = 60$ K [8]. In the presence of magnetic field $H = 14$ T, one can clearly see the "exposed" background from the normal-state pseudogap [8].

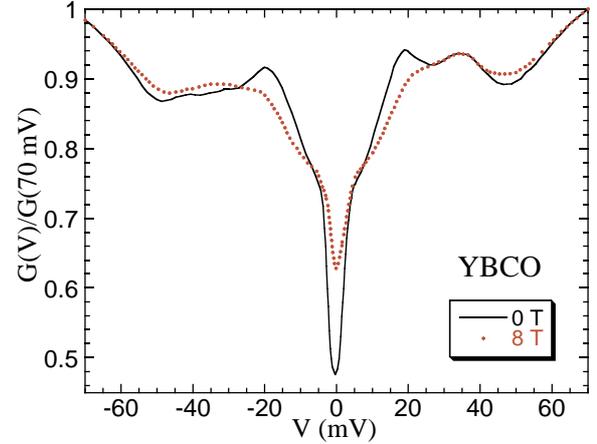

Fig. 3. SIN conductances $G(V)$ for YBCO/Pb junction in 0 T (solid line) and 8 T magnetic field applied along the $c$-axis (dotted line) at $T = 10$ K [9].

The difference in origin between the normal-state background and the QP peaks can be seen in tunneling conductances obtained not only in Bi2212 but in $YBa_2Cu_3O_{7-\delta}$ (YBCO) as well. Figure 3 depicts two condactances measured in YBCO/Pb junction at 10 K [9]. The dotted curve is obtained in magnetic field of 8 T applied along the $c$-axis. In Fig. 3, one can see at positive bias that the magnetic field suppresses the QP peak, while the hump remains unchanged.

To summarize, on the basis of experimental data obtained in Bi2212 and YBCO, it was shown that the dips in tunneling conductances appear naturally in the spectra due to a superposition of the peaks and humps and, therefore, have no physical meaning.

**References**

[1] J. G. Bednorz and K. A. Müller, Z. Phys. B 64 (1986) 189.
[2] T. Takahashi, J. Phys. Chem. Sol. 67 (2006) 193.
[3] Y. DeWilde, N. Miyakawa, P. Guptasarma, M. Iavarone, I. Ozyuzer, J. F. Zasadzinski, P. Romano, D. G. Hinks, C. Kendziora, G. W. Grabtree, and K. E. Gray, Phys. Rev. Lett. 80 (1998) 153.
[4] A. Mourachkine, Europhys. Lett. 55 (2001) 559.
[5] A. Mourachkine, Mod. Phys. Lett. B 19 (2005) 743.
[6] A. Mourachkine High-Temperature Superconductivity in Cuprates: The Nonlinear Mechanism and Tunneling Measurements, Kluwer Academic, Dordrecht, 2002, pp. 145, 60.
[7] D. L. Feng, D. H. Lu, K. M. Shen, C. Kim, H. Eisaki, A. Damascelli, R. Yoshizaki, J.-i. Shimoyama, K. Kishio, G. D. Gu, S. Oh, A. Andrus, J. O'Donnell, J. N. Eckstein, and Z. X. Sheen, Science 289 (2000) 277.
[8] V. M. Krasnov, A. E. Kovalev, A. Yurgens, and D. Winkler, Phys. Rev. Lett. 86 (2001) 2657.
[9] M. Gurvitch, J. M. Valles Jr., A. M. Cucolo, R. C. Dynes, J. P. Garno, L. F. Schneemeyer, and J. V. Waszczak, Phys. Rev. Lett. 63 (1988) 1008.